# Performance Analysis and Optimization of Lumped Parameters of Electrostatic Actuators for Optical MEMS Switches

D.Mohana Geetha
Department of Electronics and Communication
Engineering
Kumaraguru college of Technology
Coimbatore641006, India.

M.Madheswaran
Center for Advanced Research
Department of Electronics and Communication
Engineering
Muthayammal Engineering College,
Rasipuram 637408 , India.

*Abstract – This Paper deals design and simulation of electro statically actuated clamped-clamped beam and cantilever beam using finite element analysis method (FEM). A detailed study and performance analysis for various bias voltages is provided in this paper. The pull in voltages for different dimensions of the beams and the natural or the dominant modes and the corresponding eigen values have been studied for different bias voltages. The displacement of the beam is also studied for various dimensions of the beam. The results were obtained for the length and the width of both clamped – clamped beam and cantilever beam through extensive simulations. The results obtained shows that pull in voltages varies from 2348V to 772V and the natural frequencies vary from 102.92 KHz to 916.35 KHz.*

## I. INTRODUCTION

The development in the field of MEMS and its application in optical domain have increased the attention of research in the recent past.MEMS are found suitable for optical applications because these devices can be matched to optical wavelengths, and manufactured in high volume and high density arrays in the semiconductor manufacturing processes [1]. The inherent advantages of MEMS have started replacing the optical transmitting switches and tunable filters with MEMS actuators [2].The most common process in all MEMS devices is actuation which affect the mechanical motion, forces, and work by a device or system on its surroundings in response to the application of a bias voltage or current. The most common types of actuators are electrostatic, thermal, magnetic, piezoelectric, shape memory alloys, and hydraulics. MEMS based devices are highly used for several applications like biomedical sensors, miniature biomedical instruments ,cardiac management systems, neuro stimulation ,engine and propulsion control, automotive safety, braking and suspension systems, telecommunication optical fiber components and switches, data storage systems, electromechanical signal processing and also for military applications. The drive mechanism of these devices includes a constant voltage source (voltage drive) or constant current source (current drive) to enable electrostatic actuation or capacitive sensing. Sazzadur Chowdhury, M. Ahmadi, W. C. Miller have demonstrated that the electrostatic MEMS devices can be implemented with less complexity[3]. Optical MEMS devices is an electro statically actuated micromechanical mirror [1].

Jin Cheng, Jiang Zhe, Xingtao Wu, K.R.Farmer, V.Modi, Lu Frechette,2004 [4] carried out the pull in analysis for cantilever beam and fixed- fixed beam and have calculated the travel ranges of free deformable actuators to 47.2% and 42% respectively. Ofir Bochobza –Degani, Eran Socher,Yael Nemirovsky [5] had modeled a general electrostatic actuator with a charge distribution in the dielectric coating and results were obtained for pull in voltages. Joseph I.Seeger and Bernhard E.boser [6] demonstrated and showed that the structure can move beyond the well known pull in limit at resonance Sazzadur Chowdhury, M. Ahmadi, W. C. Miller provided the comparative study of closed form methods for calculating the pull in voltage of electro statically actuated fixed- fixed beam actuators[3 ].

The electrostatic force associated with the constant voltage drive mode becomes nonlinear and gives rise to the well-known phenomenon of 'pull-in'. The pull-in phenomenon causes an electro statically actuated beam to collapse on the ground plane if the drive voltage exceeds certain limit depending on the device geometry. Accurate determination of the pull-in voltage is critical in the design process to determine the sensitivity, frequency response, instability, distortion, and the dynamic range of the device [3]. Since the determination of accurate pull in voltage and natural frequency is essential, extensive simulation is carried out by varying the dimensions of clamped –clamped beam and cantilever beam.

## II. MODELING OF ELECTROSTATIC ACTUATORS

Electrostatic actuators have fast response and low power consumption. Application and release of forces take virtually the same time which is not the case of thermo actuation because of fast heating and slow cooling. Electro statically driven actuators are less sensitive to environmental conditions than others. They are two metal structures separated by an air gap. A bias voltage is applied between the metal structures, which results in a separation of charges between them. This produces an electrostatic force that can be used to decrease the gap between the plates as shown in the Fig 1.

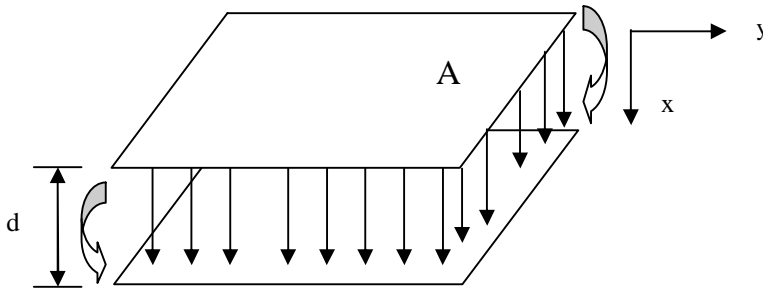

Figure.1.Schmatic structure of parallel plate capacitor

The two plates have an overlapping area of A and a spacing of d. The dielectric constant or relative electrical permittivity of the media between the two plates is denoted $\varepsilon_r$ .The permittivity of the media is $\varepsilon = \varepsilon_r \; \varepsilon_o$,where $\varepsilon_o$ is the permittivity of the vacuum[7].

The value of the capacitance C, between the two parallel plates is given by

$$C = \frac{Q}{V} \qquad (1)$$

where Q is the amount of stored charge and V is the electrostatic potential.

The electric energy stored in the capacitor is given by

$$U = \frac{1}{2} C V^2 = \frac{1}{2} \frac{Q^2}{C} \qquad (2)$$

According to Gauss's law, the magnitude of the primary electric field E is given by

$$E = \frac{Q}{\varepsilon A} \qquad (3)$$

The magnitude of the voltage is the electric field times the distance between two plates 'd'.

The capacitance of the parallel plate capacitor is

$$C = \frac{Q}{V} = \frac{Q}{E.d} = \frac{Q}{\frac{Q}{\varepsilon A} d} = \frac{\varepsilon A}{d} \qquad (4)$$

The capacitor can be used as an actuator to generate force or displacement .As a differential voltage is applied between the two parallel plates, an electrostatic attraction force is developed. The magnitude of the forces equals the gradient of the stored electric energy $V_s$ with respect to the dimensional variable .The magnitude of the force is

$$F = \left| \frac{\partial U}{\partial x} \right| = \frac{1}{2} \left| \frac{\partial C}{\partial x} \right| V^2 \qquad (5)$$

where x is the dimensional variable.

If the plate moves, the gap between the plates changes and the magnitude of the force can be given as

$$F = \left| \frac{\partial U}{\partial x} \right| = \frac{1}{2} \frac{\varepsilon A}{d^2} V^2 = \frac{1}{2} \frac{C V^2}{d} \qquad (6)$$

with normal dimension changed from x to d.

The suspended plate is attracted towards the bottom plate due to the resultant electrostatic force. The suspended plate move towards the bottom plate until an equilibrium exits between them. The suspended plate makes contact with the bottom plate when there is maximum electrostatic force.

The capacitance of the device over a range of motion can be used to characterize the electromechanical response of the device.

An electrostatic actuator can be modeled as a variable capacitor suspended by elastic springs. An important design aspect for electrostatic actuators is to determine the amount of static displacement under a certain biased voltage as shown Fig 2. The upper beam is supported by a mechanical spring with a force constant $K_m$ .Gravitational force can be neglected because the mass of the beams are very small.

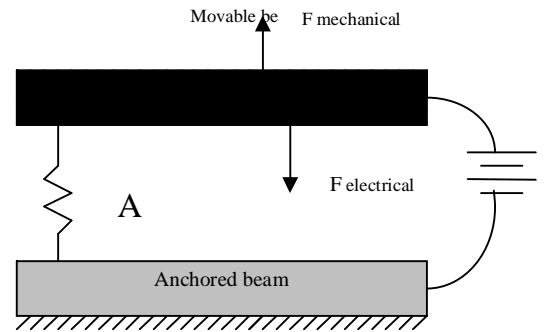

Figure.2.A coupled electromechanical model

When a voltage is applied, an electrostatic force $F_{electric}$ is developed. The magnitude of $F_{electric}$ is given by

$$F_{electric} = \frac{1}{2} \frac{c V^2}{d} \qquad (7)$$

This force will tend to decrease the gap and gives displacement and mechanical restoring force. Under static equilibrium, the mechanical restoring force has an equal magnitude but opposite direction as the electrostatic force.

The electrostatic force modifies the spring constant and the spring will be softened. The spatial gradient of the electric force is defined as an electrical spring constant and given by

$$K_e = \left| \frac{\partial F_{electric}}{\partial d} \right| = \left| -\left( \frac{cv^2}{d^2} \right) \right| = \frac{cV^2}{d^2}. \tag{8}$$

The electrostatic force at equilibrium when the beam is applied with a bias voltage is given by

$$F_{electric} = \frac{1}{2} \varepsilon A V^2 (x_0 + x)^2 = \frac{\frac{1}{2} c(x) V^2}{x_0 + x} \tag{9}$$

The magnitude of the mechanical restoring force is given by

$$F_{mechanical} = -K_m x \tag{10}$$

Equating the magnitudes of $F_{mechanical}$ and $F_{electric}$ at x and rearranging the terms, the displacement can be calculated as

$$-x = \frac{F_{mechanical}}{K_m} = \frac{F_{electric}}{K_m} = \frac{c(x) V^2}{2 (x_{x=0}) K_m} \tag{11}$$

At a particular bias voltage, mechanical restoring force and the electrostatic force balance each other. The magnitude of electric force constant equals the mechanical force constant. The effective force constant of the spring is zero. The bias voltage invokes such a condition is called the pull in voltage 'V$_p$'. If the bias voltage is increased beyond V$_p$, the equilibrium position disappears. The electrostatic force continues to rise while the mechanical force increases linearly only. The two beams are pulled against each other and they make contact. This condition is called pull in or snaps in.

At the pull in condition the magnitudes of electrical force and mechanical force and can be equated as

$$V^2 = -\frac{2 K_m x (x + x_0)^2}{\varepsilon A} = -\frac{2 K_m x (x + x_0)^2}{c} \tag{12}$$

The value of x is negative when the spacing between two electrodes decreases. The gradients of these two forces at the intersection point are equal and given by

$$\left| K_e \right| = \left| K_m \right| \tag{13}$$

From equation (8) and (12)

$$K_e = \frac{cV^2}{(x + x_0)^2} = -2 K_m x (x + x_0) \tag{14}$$

The solution of x can be obtained as $x = -\dfrac{x_0}{3}$ (15)

From the equation (15), it may be concluded that the relative displacement of the parallel plate from its initial position is exactly one third of the original spacing at the critical pull-in voltage.

The voltage at the pull in can be estimated from (12) and equation (15).

The pull in voltage is given by

$$v_p = \frac{2 x_0}{3} \sqrt{\frac{K_m}{1.5 c}} \tag{16}$$

### III. COMPUTATIONAL TECHNIQUE

The clamped-clamped beam and cantilever beam actuators can be simulated and the performance in the electrical domain can be analyzed using Finite element method. The flow graph of the simulation is shown in Fig 3. The structure of the clamped – clamped and cantilever beam actuators are shown in Fig 4. The pull in voltage and the natural frequency of the proposed actuators can be simulated using ANSYS. The region between the two plates can be considered as the active region and the capacitance can be estimated by dividing the beam into meshes. Reduced order modeling of the coupled electrostatic system can be used to replace the electrostatic mesh for improving the execution time.

The eigen value and the eigen vectors can be calculated using

$$[K]\{\Phi_i\} = \lambda_i [M]\{\Phi_i\} \tag{18}$$

where [K] = structure stiffness matrix

$\{\Phi_i\}$ = eigenvector

$\lambda_i$ = eigen value

[M] = structure mass matrix

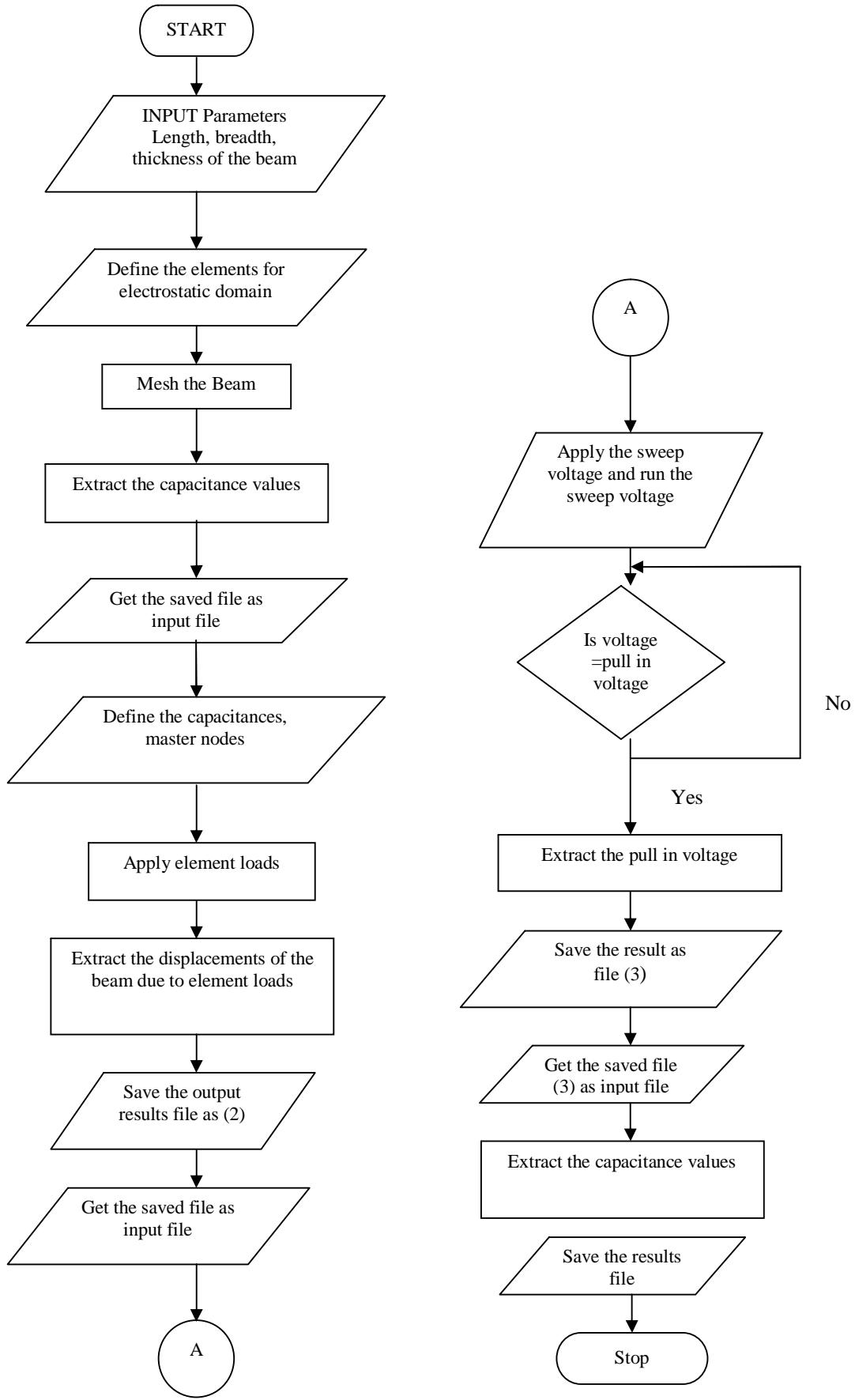

**Figuer.3.Flow chart for Pull in voltage and extracting capacitance**

The structure of the stiffness and mass matrix can be reduced by considering the set of degrees of freedoms (DOFs).Then using Guyan reduction technique the relationship can be rewritten as $\left[ k \right]\left\{ \hat{\phi}_i \right\} = \lambda_i \left[ \hat{M} \right]\left\{ \hat{\phi}_i \right\}$  (19)

Where $\left[ k \right]$ = reduced stiffness matrix (known)

$\left\{ \hat{\phi}_i \right\}$ = eigenvector (unknown)

$\lambda_i$ = eigen value (unknown)

$\left[ \hat{M} \right]$ = reduced mass matrix (known)

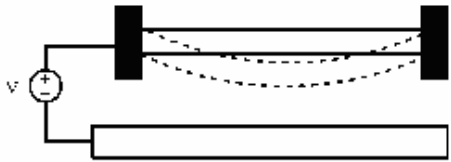

Figure 4(a).Clamped-Clamped beam

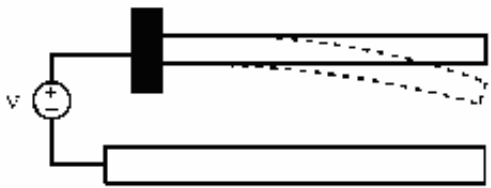

Figure 4(b) Cantilever beam

Then the actual eigen value can be extracted using HBI (Householder-Bisection-Inverse iteration).Using FEM and the ANSYS , the pull in voltage and the natural frequency can be obtained to estimate the performance of both the actuators to find their capacity to use them in optical MEMS switches.

## V. RESULTS AND DISCUSSION

The electrostatic actuators with clamped-clamped beam and the cantilever beam structures has been  simulated using ANSYS and the parameters have been estimated using FEM analysis.The simulated output  of the clamped-clamped  beam and cantilever electrostatic actuators is shown in Fig 5(a) and 5(b)respectively. The pull in voltage for various beam length

has been  calculated for different beam width of clamped-clamped beam actuators  and shown in Fig .6

It is seen that the pull in voltage decreases with beam length as well as beam width. However it is clear from the results that the variation in pull in voltage is much significant for the lower beam width. The pull voltage for cantilever beam with various beam width and beam length is shown in Fig.7.It is seen from the Fig that the pull in voltage for the beam length 90 μm and 2 μm beam width is 450V.It is found that at the same beam width and the beam length the pull voltage was estimated as 1100v in clamped-clamped beam actuators. The natural frequency for both clamped –clamped beam and cantilever beam actuators are shown in Table.I

It is seen from the table that the natural frequency is inversely proportional to the beam length in both the cases. The variation of natural frequency with the length and width of the cantilever beam is shown in the Fig .8

The displacement of the beam for length L=85 μm for both clamped –clamped and cantilever beam in Fig.9.It is seen from the Fig that the displacement is much significant in cantilever beam for the beam with less width.

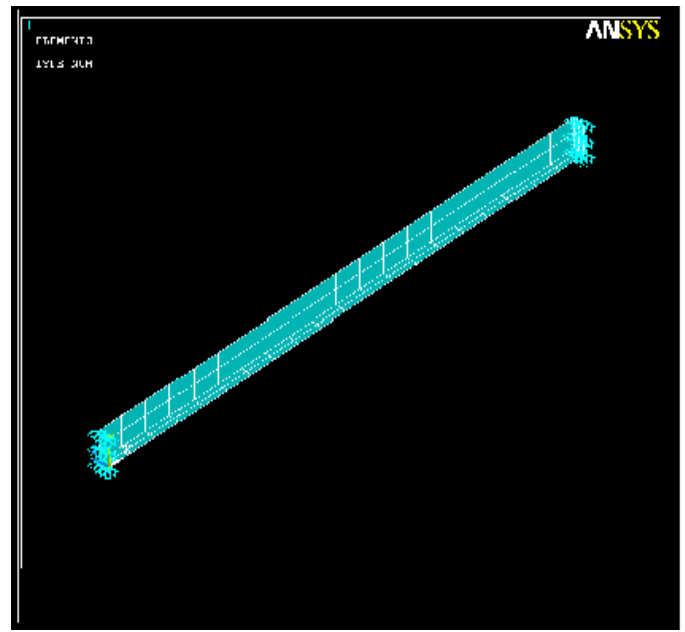

**Figure 5(a).Output of displaced clamped-clamped beam**.

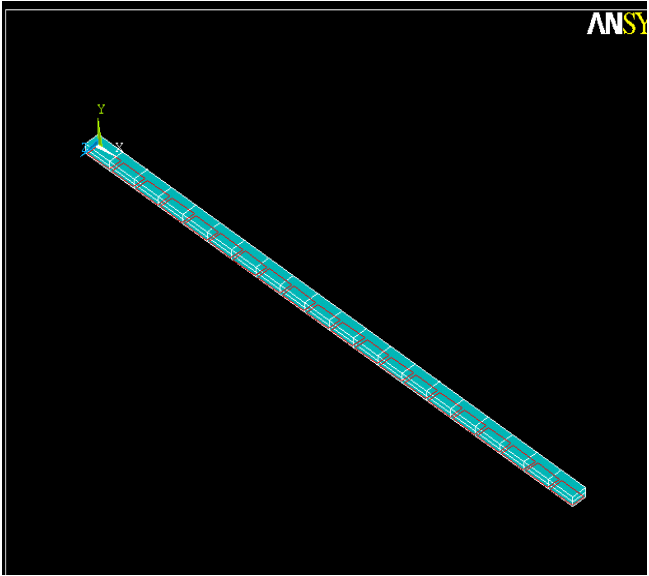

**Figure 5(b).Output of displaced cantilever beam**.

NATURAL FREQUENCIES FOR CLAMPED AND CANTILEVER BEAMS

| Beam Width in µm | Beam Length in µm | Natural Frequencies (Hz) | |
|---|---|---|---|
| | | Clamped-clamped beam | Cantilever beam |
| 15 | 100 | 0.10292E+08 | 0.48965E+07 |
| 10 | 100 | 0.10289E+08 | 0.48960E+07 |
| 5 | 100 | 0.10285E+08 | 0.48957E+07 |
| 2 | 100 | 0.10284E+08 | 0.17325E+07 |
| 1 | 100 | 0.10284E+08 | 0.17325E+07 |
| 15 | 95 | 0.10294E+08 | 0.54232E+07 |
| 10 | 95 | 0.11337E+08 | 0.54226E+07 |
| 5 | 95 | 0.11333E+08 | 0.54222E+07 |
| 2 | 95 | 0.11332E+08 | 0.19193E+07 |
| 1 | 95 | 0.11332E+08 | 0.19193E+07 |
| 15 | 90 | 0.12566E+08 | 0.60395E+07 |
| 10 | 90 | 0.12562E+08 | 0.60389E+07 |
| 5 | 90 | 0.12558E+08 | 0.60384E+07 |
| 2 | 90 | 0.12556E+08 | 0.21381E+07 |
| 1 | 90 | 0.12556E+08 | 0.21381E+07 |
| 15 | 85 | 0.14011E+08 | 0.67670E+07 |
| 10 | 85 | 0.14006E+08 | 0.67662E+07 |
| 5 | 85 | 0.14001E+08 | 0.67657E+07 |
| 2 | 85 | 0.14000E+08 | 0.23965E+07 |
| 1 | 85 | 0.13999E+08 | 0.23965E+07 |
| 15 | 80 | 0.15730E+08 | 0.76340E+07 |
| 10 | 80 | 0.15725E+08 | 0.76330E+07 |
| 5 | 80 | 0.15719E+08 | 0.76324E+07 |
| 2 | 80 | 0.15717E+08 | 0.27046E+07 |
| 1 | 80 | 0.15716E+08 | 0.27046E+07 |
| 15 | 75 | 0.17797E+08 | 0.86784E+07 |
| 10 | 75 | 0.17791E+08 | 0.86773E+07 |
| 5 | 75 | 0.91649E+07 | 0.86765E+07 |
| 2 | 75 | 0.17781E+08 | 0.30762E+07 |
| 1 | 75 | 0.17781E+08 | 0.30762E+07 |

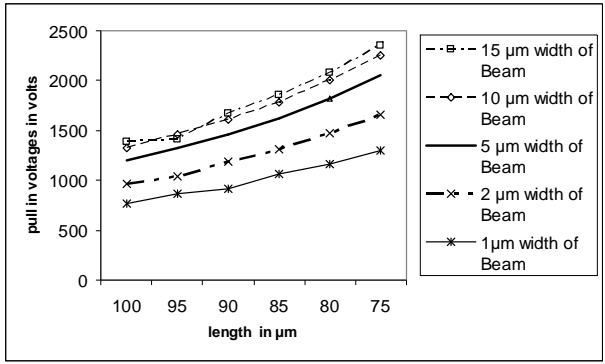

**Figure. 6 Pull in voltages  of clamped-clamped**

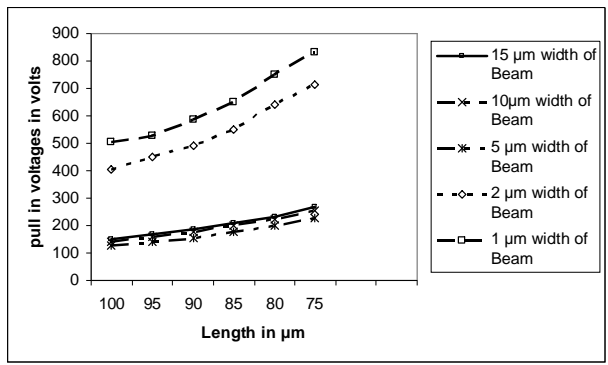

**Figure.7  Pull in voltage of cantilever beam.**

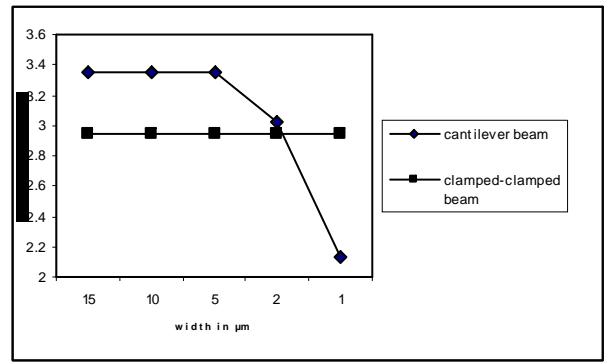

**Figure.9 Variation of displacement of the beams**
**for length L=85 µm.**

## VI. CONCLUSION

The pull in voltages calculations was made using finite element method. ANSYS with Reduced order modeling was used to extract results quickly. Fringing field effects are also considered to yield accurate results. Even with the strongly coupled lumped transducers, convergence issues were experienced when applied to the difficult hysteric pull-in and release analyses. The cause of the problem can be attributed to the negative total system stiffness matrix and can be resolved using the augmented stiffness method.

## AUTHORS PROFILE

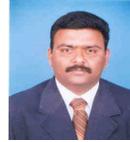
Dr. M. Madheswaran has obtained his Ph.D. degree in Electronics Engineering from Institute of Technology, Banaras Hindu University, Varanasi in 1999 and M.E degree in Microwave Engineering from Birla Institute of Technology, Ranchi, India. He has started his teaching profession in the year 1991 to serve his parent Institution Mohd. Sathak Engineering College, Kilakarai where he obtained his Bachelor Degree in ECE. He has served KSR college of Technology from 1999 to 2001 and PSNA College of Engineering and Technology, Dindigul from 2001 to 2006. He has been awarded Young Scientist Fellowship by the Tamil Nadu State Council for Science and Technology and Senior Research Fellowship by Council for Scientific and Industrial Research, New Delhi in the year 1994 and 1996 respectively. His research project entitled "Analysis and simulation of OEIC receivers for tera optical networks" has been funded by the SERC Division, Department of Science and Technology, Ministry of Information Technology under the Fast track proposal for Young Scientist in 2004. He has published 120 research papers in International and National Journals as well as conferences. He has been the IEEE student branch counselor at Mohamed Sathak Engineering College, Kilakarai during 1993-1998 and PSNA College of Engineering and Technology, Dindigul during 2003-2006. He has been awarded Best Citizen of India award in the year 2005 and his name is included in the Marquis Who's Who in Science and Engineering, 2006-2007 which distinguishes him as one of the leading professionals in the world. His field of interest includes semiconductor devices, microwave electronics, optoelectronics and signal processing. He is a member of IEEE, SPIE, IETE, ISTE, VLSI Society of India and Institution of Engineers (India).

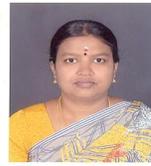
Mrs.D.Mohanageetha obtained her B.E degree from Bharathiar University in 1996, M.E Degree in Communication systems from Madurai Kamaraj University, Madurai in 2000. She has started her teaching profession in the year 1996 at V.L.B Janakiammal College of Engineering and Technology, Coimbatore. At present ,she is an Assistant Professor in Department of Electronics and Communication, Kumaraguru college of Technology, Coimbatore, Tamil nadu,India . She has published 10 research papers in International and national conferences. She is a part time Ph.D research scalar in Anna University Chennai. Her areas of interest are Optical networking,, MEMS,EMI/EMC and Image processing. She is a life member of ISTE and member of IEEE.